\documentclass[conference]{IEEEtran}
\usepackage{cite}
\usepackage{amsmath,amssymb,amsfonts}
\usepackage{algorithmic}
\usepackage{graphicx}
\usepackage{textcomp}
\usepackage{ascmac}
\usepackage{xcolor}
\usepackage{lscape}
\usepackage{listings}
\usepackage{float}

\def\BibTeX{{\rm B\kern-.05em{\sc i\kern-.025em b}\kern-.08em
    T\kern-.1667em\lower.7ex\hbox{E}\kern-.125emX}}
\begin{document}

\title{An Empirical Analysis of Git Commit Logs \\for Potential Inconsistency in Code Clones}

\author{\IEEEauthorblockN{Reishi Yokomori}
\IEEEauthorblockA{\textit{Dept. of Software Engineering} \\
\textit{Nanzan University}\\
Nagoya, Japan\\
yokomori@nanzan-u.ac.jp}
\and
\IEEEauthorblockN{Katsuro Inoue}
\IEEEauthorblockA{\textit{Dept. of Software Engineering}\\
\textit{Nanzan University}\\
Nagoya, Japan\\
inoue599@nanzan-u.ac.jp}
}

\maketitle

\begin{abstract}

Code clones are code snippets that are identical or similar to other snippets within the same or different files. They are often created through copy-and-paste practices and modified during development and maintenance activities.
Since a pair of code clones, known as a clone pair, has a possible logical coupling between them, it is expected that changes to each snippet are made simultaneously (co-changed) and consistently.
There is extensive research on code clones, including studies related to the co-change of clones; however, detailed analysis of commit logs for code clone pairs has been limited.

In this paper, we investigate the commit logs of code snippets from clone pairs, using the git-log command to extract changes to cloned code snippets. We analyzed 45 repositories owned by the Apache Software Foundation on GitHub and addressed three research questions regarding
commit frequency, co-change ratio, and commit patterns. 
Our findings indicate that (1) on average, clone snippets are changed infrequently, typically only two or three times throughout their lifetime, (2) the ratio of co-changes is about half of all clone changes, with 10-20\% of co-changed commits being concerning (potentially inconsistent), and (3) 35-65\% of all clone pairs being classified as concerning clone pairs (potentially inconsistent clone pairs).
These results suggest the need for a consistent management system through the commit timeline of clones.

\end{abstract}


\section{Introduction}

A code clone is a code snippet that has identical or similar counterparts in the same or different files. It is sometimes referred to as a clone or duplicated code\cite{roy07,inoue21-2}. Code clones are created during software development and maintenance, either intentionally or unintentionally, and are often considered a bad smell in code\cite{fowler99, geiger06}. 
Code clones can decrease program quality by increasing program length and can also reduce maintainability by requiring consistent changes across the clones\cite{betterburg09, jens07}.

Numerous studies on code clones have been conducted \cite{inoue21, koschke07, rattan13, roy09}, and several have focused on refactoring and removing clones \cite{higo04,higo08,mondal20}. However, due to constraints related to programming languages or system design, not all clones can be removed or are advised to be removed\cite{kapser06}. Instead, they are often left unchanged, or sometimes changed and evolved, with or without proper management\cite{kim05,mostafa19,bladel23}.

Code clones can create a logical coupling between cloned snippets, necessitating consistent changes. In many cases, if one snippet of a clone pair is modified, the corresponding snippet in the pair should also be modified in the same way.  One indicator of consistent changes is the simultaneous modification of clone snippets (co-changed). However, it is not well known how frequently code clones are changed and how many of these changes are co-changes during the development and maintenance of software systems. Additionally, it is important to understand whether co-changed clones are always consistent and what change patterns might lead to inconsistencies. These issues have not been thoroughly investigated so far.

We are interested in the details of code clone changes, such as their frequency and patterns. The purpose of this study is to empirically investigate the characteristics of code clone changes. Based on the results, we aim to identify patterns that pose higher risks and contribute to the development of safer code clone management methods.

In this paper, we have established three research questions to reveal the characteristics of code clone changes. Based on these research questions, we conducted a detailed analysis of 45 Apache repositories using Git commit logs as the evolution history of clone changes.

Our findings indicate that, on average, clone snippets are changed infrequently, typically only two or three times throughout their lifetime. Additionally, the ratio of co-changes is about half of all clone changes, with 10-20\% of co-changed commits being concerning (potentially inconsistent commits), and 35-65\% of all clone pairs being classified as concerning clone pairs (potentially inconsistent clone pairs).

Contributions of this paper are as follows.

\begin{itemize}

\item We have devised a method for analyzing Git commit logs to characterize the changes in code clone pairs.

\item We have developed a method for identifying concerning clone pairs using the similarity of patches (diffs) in the commit logs.

\item Using these methods, we performed an empirical analysis of 45 typical open source software projects from Apache repositories, providing statistics and characteristics of clone commits.

\item Based on the analysis results, we have presented implications for the features of clone management tools.

\end{itemize}

The paper is organized as follows. Section II describes the terms and their definitions. Section III discusses related works. We present three research questions in Section IV, and the research approach in Section V. Section VI presents the analysis results. Section VII discusses the results and their implications for clone management. Section VIII addresses the threats to the validity of this study. We conclude with future work in Section IX.

\section{Terms and Definitions}

A \emph{code snippet}, or simply a \emph{snippet}, is a part of a source code within a software system. Sometimes, we duplicate a code snippet by copying and pasting it, then modify the pasted code with the different variable names or literals,  within the same software system or across different ones.

A pair of two code snippets that are the same or similar is called a \emph{clone pair}, and 
each snippet of the pair is referred to as a \emph{clone}, \emph{code clone}, or \emph{(code) clone snippet}\cite{inoue21-2,roy07}. 
`Clone' means a code snippet having another code snippet (called \emph{partner}) of a clone pair within the same file or different files in the same project repository. We do not consider here inter-project clones.

The \emph{length} of a clone (or \emph{clone length}) is the lines of code (LOC) of the clone, which may
include non-executable lines such as comments or blank lines.
A \emph{clone set} (sometimes called clone class or clone group) is a set of code snippets whose any two elements form a clone pair. 
The \emph{size} of a clone set is the number of elements (snippets) of the clone set.
%

A clone pair is generally categorized by its similarity levels as follows\cite{carter1993clone,roy07,inoue21-2}.

\emph{Type-1} clone pair is syntactically identical code snippets, with possible differences of non-executable elements such as white spaces, tabs, comments, and so on.
\emph{Type-2} pair is structurally identical code snippets, with possible differences of identifier names, literals, and type names, in addition to the differences of type-1.
\emph{Type-3} pair is similar code snippets, with possible differences of several statements added, removed, or modified to another snippet, in addition to the differences of type-2. 
\emph{Type-4} pair is syntactically different code snippets, but their functionalities are the same.

Here, we mainly discuss type-1 and type-2 clones, in addition to type-3 clones, which are handled as the composition of smaller type-1 or type-2 clones. Type-4 clones are excluded from the scope of this study because they often require complex modifications from the original code and are unlikely to result from straightforward copy-and-paste actions\cite{roy09}.



A \emph{commit} in Git is an action to save all changes on the code to the Git repository, and a \emph{co-changed commit} is a commit of changes (modifications) made to two snippets forming a clone pair. The \emph{commit log} in Git is a record that lists all the commits made to the repository or a code snippet, showing the metadata composed of commit hashes, timestamps, authors, messages, and code diff (patch), which helps in tracking changes and understanding the history of a project.
An \emph{inconsistent} change to a clone pair A and B means that changes to A, B, or both are not aligned with the contexts of each other, potentially introducing defects into A or B. We will show an example in Section \ref{sec:result-rq2} with Listing \ref{lst:Patch-inconsistent}.

\begin{figure*}[ht]
\centerline{\includegraphics[width=14.0cm]{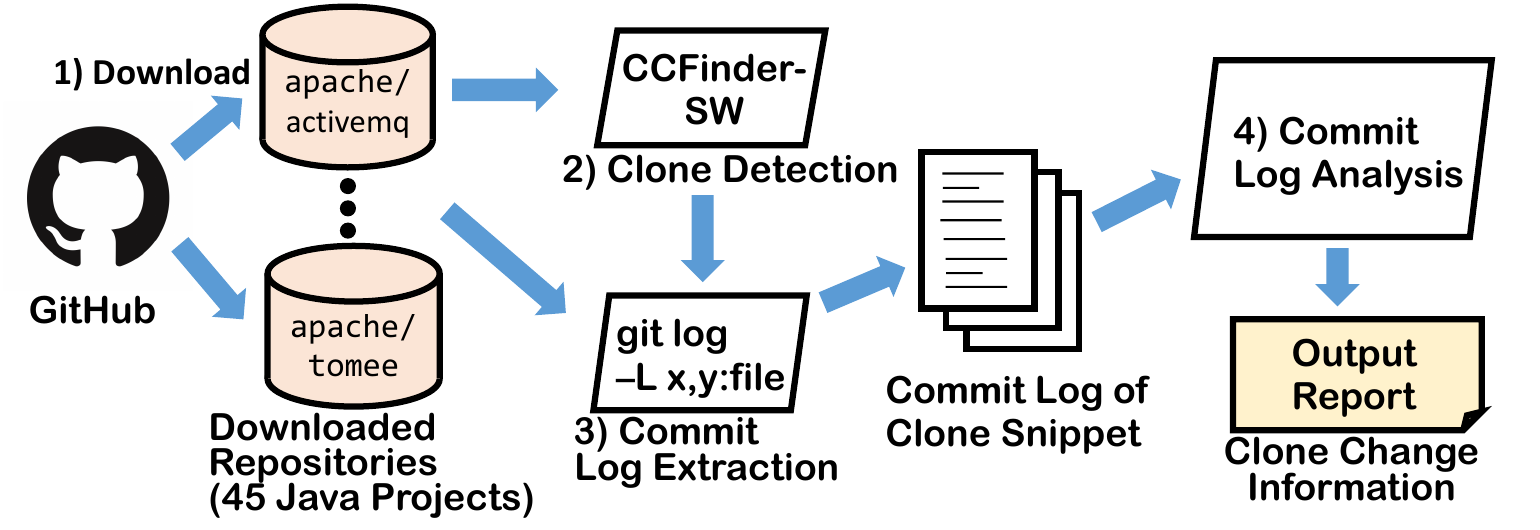}}
\caption{Overview of Analysis Method}
\label{fig:overview}
\end{figure*}

\section{Related Works}

\subsection{Code Clone Analysis}

Numerous studies on code clones have been conducted and published\cite{koschke07,roy07,inoue21,rattan13}. Initial research papers were released on methods for detecting code clones in the 1990s, and the fundamental concepts behind them were presented\cite{Baker92aprogram,carter1993clone,baxter98}. Since then, various detection techniques and tools have been developed and introduced \cite{Kamiya2002CCFinder,Cordy2011NiCad,Sajnani2016SourcererCC}. Research is actively being pursued in detection methods utilizing machine learning and type-4 clone detection\cite{OREO-FSE2018,Zhao2018DeepSim,arshad22,rabbani22}.

\subsection{Empirical Studies on Code Clones}

Numerous empirical studies have focused on code clones. 
Roy et al. presented an empirical study of function-level clone analysis using NiCAD on over 15 open-source C and Java systems, verifying detected clones and providing a comprehensive online catalog for reference and benchmarking\cite{roy08}. For fundamental statistics of empirical data on clones, Goon et al. presented the code clone ratios in C and C++ programs within open-source projects over time \cite{goon17}. Honda et al. investigated the changes in the code clone ratio from the initial development phases to the final phases and found that approximately 20\% of all source code consists of clones \cite{honda14}.

 For the empirical study of clones and bugs, Rahman et al. analyzed a significant relationship between the stability and bug-proneness of clones, with buggy clones changing more frequently than non-buggy ones\cite{rahman17}. Islam et al. found that up to 10\% of code clones may contain replicated bugs, with type-2 and type-3 clones showing higher tendencies for bug replication compared to type-1 clones\cite{islam16}.
 Hotta et al. investigated the change frequency of cloned and non-cloned code, and an experiment on 15 open-source software systems showed that duplicate code does not negatively impact software evolution\cite{hotta10}.


\subsection{Evolution and Co-Change of Code Clones}

For the evolution and genealogy of code clones, several preceding works have been presented. Kim et al. conducted a pioneering study, finding that many clones are short-lived, making extensive refactoring potentially unnecessary and that long-lived clones often resist refactoring due to language limitations\cite{kim05}. Following this work, empirical studies on code clone evolution have examined migration and mutation, fault-proneness, test and production code, lifetime, and other aspects\cite{xie14, bladel23, mostafa19, ripon10, mondal13}.

For co-change prediction of clones, Mondal et al. studied the evolutionary coupling of clone fragments with non-clone fragments and clone fragments\cite{mondal14,mondal14-2}. Their analysis granularity is function level and detailed analysis for code snippets with commit history has not yet been made. For a detailed analysis of code snippets, Modal et al. also investigated micro clones, and proposed a composite ranking mechanism based on file proximity and evolutionary coupling, improving the prediction of co-change candidates\cite{mondal19}. They focused on small and type-1 clones without chasing their history, but ours contains type-1 to 3 clones with the history analysis of the commit logs.

Marks et al. studied factors affecting the co-change frequency of code clones, revealing that reducing code complexity improves clone maintenance while increased complexity decreases co-change frequency potentially causing defects\cite{marks13}. 
In \cite{lozano14}, file-level co-changes of clones were studied, revealing that most cloned files co-change only with file-sharing clones, suggesting that consistent changes might be the norm in cloned code.
For the characteristics of co-changed clones, Yudha et al. revealed a specific length of co-changed clones with less existence of control statements\cite{yudha12}.
For inconsistent clone changes, Jens reported the ratio of inconsistent change of clone sets\cite{jens07}. Betterburg et al. reported that only a small percentage of these changes introduce defects in the system\cite{betterburg09}. 
Our results partially align with these results, although we have further investigated co-changes in code snippet level and identified potential inconsistent commits and clone pairs as concerning co-changed commits and clone pairs.

\section{Research Questions}

First, we have set up a research question to investigate the basic characteristics of the commit log for code snippets in code clone pairs.
We will investigate the stability of clones in the history of development. 

\begin{screen}[4]
\textbf{RQ1: How frequently are code clone snippets modified during development?}
\end{screen}

Next, we are interested in the co-changed commits in the commit logs of clones.

\begin{screen}[4]
\textbf{RQ2: What proportion of commits are co-changed and are the co-changed commits consistent?}
\end{screen}
We will analyze the ratio of co-changes in the commit logs and investigate the differences in source code introduced by these co-changed commits as a preliminary consistency check. While a more rigorous consistency check could be achieved by examining the correspondence of changed identifiers and literals, this requires significant effort, such as parsing sentences and matching identifiers. Therefore, this study adopts a simpler and more efficient method, as will be detailed later.




Lastly, we will investigate the patterns of concerning clone pairs in the current latest repositories.

\begin{screen}[4]
\textbf{RQ3: What are the concerning clone pairs that may need attention in the latest repository, and how prevalent are they?}
\end{screen}
This analysis leads the discussion on safe and effective clone management.

\section{Research Approach}

\subsection{Overview}

Figure \ref{fig:overview} is an overview of our analysis method, described in the following.

\begin{enumerate}
    \item We have downloaded the target repositories of Java-based Apache projects from GitHub. The details of those repositories are mentioned in the following section \ref{sec:target}. The following steps are executed for each repository.

    \item Clone detection is performed by CCFinderSW\cite{semura17}, which  is a token-based type-1 and -2 clone detector. Type-3 clones are identified as a composition of smaller type-1 or -2 clones by this tool. CCFinderSW was chosen because it is written only in Java with high portability and performance. We ran it for the source code of the latest commits, with the following parameters: minToken=50 (the minimum number of tokens to detect), rnr=0.8 (the minimum rate of non-repeating parts in the clone fragments), and tks=12 (the minimum number of token types in the clone fragments). These parameters have been chosen to filter out smaller and non-intentional similar code snippets from the output.
    The output of CCFinderSW consists of the target file information and the clone set information, both of which are used for the following steps.

    \item \texttt{git log -L x,y:file} commands are executed for all obtained clone pairs. Each of these commands generates the commit log for the code snippet between lines $x$ and $y$ in the specified file. A commit log is composed of one or more pieces of commit information, which include the hash, timestamp, author, commit message, and diff (patch).
    We specify the snippets of clone pairs obtained from the previous step to generate the change history, i.e., the evolution of the clone snippets\footnote{We have not investigated whether the snippet pairs from the middle of their history make clone pairs, as even non-clone snippet pairs require consistent changes to maintain the consistency for the latest clone snippets.}.

    \item Each git-log output is decomposed into single commit information, and the hash, timestamp, and patch are extracted for further analysis of the research questions. Also, various statistics are measured and reported as the output.

\end{enumerate}

After completing step 1), the total execution of steps 2)-4) to analyze all 45 repositories took almost a half day on M1 MacBook Pro with 16GB memory. For analysis of only Ant repository on the same machine, the execution time for Step 2) required 42.4 Sec.,  Step 3) used 482.2 Sec., and Step 4) consumed 55.0 Sec.

\subsection{Target Selection}
\label{sec:target}

We have chosen the Java-based Apache project's repositories on GitHub as the target for our study.
Apache projects are generally active and well-maintained over the years. This would be an important basis for extracting typical characteristics of commit logs of code clones.
Also, the Apache projects on GitHub encompass a diverse range of repositories, varying in size from small to large with a small to large number of contributors.

We downloaded 45 Apache-owned project repositories on GitHub on March 23, 2024, meeting criteria including repository sizes between 50MB and 100MB, more than 100 stars, and written in Java programming language\footnote{Those are activemq, activemq-artemis, ant, apex-malhar, archiva, aries, atlas, calcite, camel-kafka-connector, cayenne, commons-compress, derby, directory-server, drill, dubbo, eventmesh, felix-dev, hama, iceberg, ignite-3, incubator-baremaps, incubator-streampark, incubator-tuweni, inlong, jackrabbit, jackrabbit-oak, jclouds, jspwiki, karaf, linkis, logging-log4j2, maven, metron, myfaces, olingo-odata4, openmeetings, orc, ozone, phoenix, pig, shenyu, stanbol, struts, syncope, and tomee.}. These constraints were used to match our computing resource limitations and to exclude projects that are not popular. 
We have used the latest Java files for the clone detection and the following commit log analysis, excluding test files that contain the term "test" in their file or path names.
Also, we have excluded clone sets whose size is more than two, i.e., clone sets with more than two clone snippets. This exclusion limits the targets to clone sets with only two snippets, but it simplifies the analysis and interpretation of the result significantly. Since it might miss some important characteristics of repeatedly copied clone snippets, we will discuss this issue later in Section \ref{sec:morethan2}.

\begin{figure}[t]
\centerline{\includegraphics[width=9.0cm]{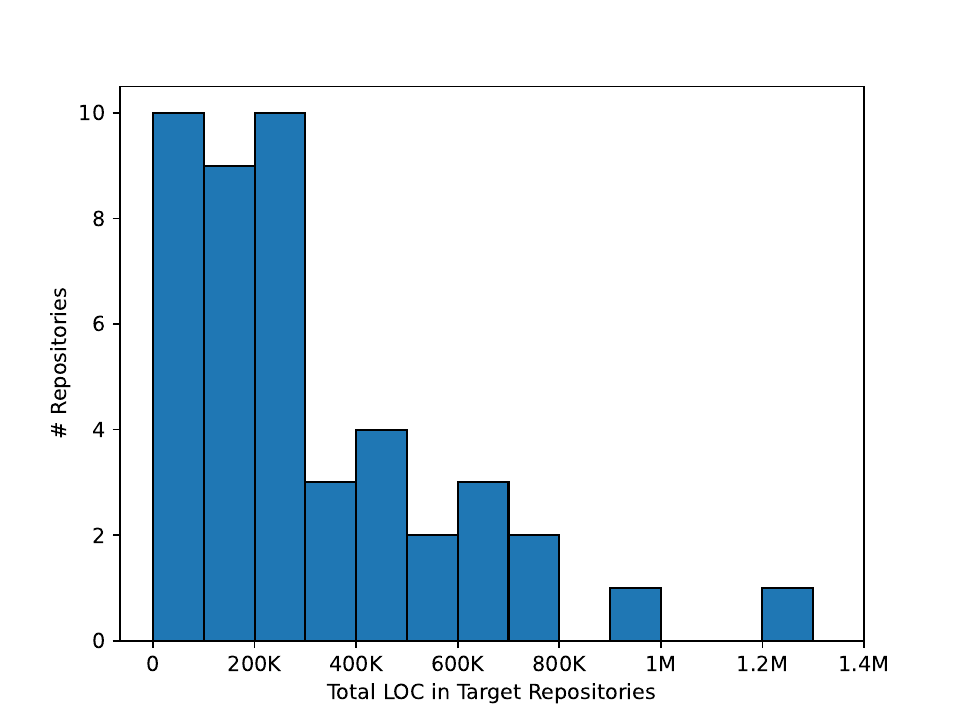}}
\caption{Distribution of the Total Lines of Code of Target Repositories}
\label{fig:Dist-loc}
\end{figure}

Table \ref{tab:all-proj} displays the characteristics of the Ant repository along with all target repositories, including Ant. Since the Ant repository is well-known and well-maintained, and its values fairly align with the medians of all repositories, we will use it for detailed explanations.
Figure \ref{fig:Dist-loc} presents the distribution of sizes in LOC (Lines Of Code), which shows that the projects are generally small, with fewer than 300K LOC. The average is 305.6K LOC, and the median is 229.0K LOC.
The number of clone sets ranges from 15 to 2,713, which is relatively small because we have filtered out simple-structured clones such as getter or setter methods using the parameter settings of the clone detector CCFinderSW.

\begin{table}[htbp]
\centering
\caption{Basic Statistics of Ant and All 45 Repositories}
\begin{tabular}{|l|r|r|}
\hline
Repositories  & Ant & All (45 incl. Ant) \\
\hline
\hline
\#Files (excl. test files) & 1,332 &270-5,873 (1,747) \\
\hline
Total K LOC & 287.6  &41.1-1,240.1 (229.0)\\
\hline
Repository age in days & 8,836  &1,034-8,836 (4,538)\\
\hline
\#Clone set & 500  & 15-2,713 (672) \\
\hline
Median clone length in LOC & 19  & 11-24 (14) \\
\hline
Total commit count &  14,954  &  1,453-19,079 (5,920) \\
\hline
\end{tabular}
\\~~~~~~~~~~~~~~~~~~~~~~ (Parentheses values are median for all)
\label{tab:all-proj}
\end{table}
\noindent\begin{minipage}{\columnwidth} 
\begin{lstlisting}[caption={Snippet-A: .../tools/ant/UnknownElement.java},label=lst:Snippet-A, firstnumber=409]
if (u.children != null) {
    List<UnknownElement> newChildren = new ArrayList<>(u.children);
    if (children != null) {
        newChildren.addAll(children);
    }
    children = newChildren;
\end{lstlisting}

\begin{lstlisting}[caption={Snippet-B: .../tools/ant/RuntimeConfigurable.java}, label=lst:Snippet-B, firstnumber=589]

// Children (this is a shadow of UnknownElement#children)
if (r.children != null) {
    List<RuntimeConfigurable> newChildren = new ArrayList<>(r.children);
    if (children != null) {
        newChildren.addAll(children);
    }
    children = newChildren;
\end{lstlisting}
\end{minipage}

\noindent\begin{minipage}{\columnwidth} 
\begin{lstlisting}[caption={Timestamps of Snippet-A},label=lst:TS-A, firstnumber=1]
Thu Nov 6 09:04:08 2003 +0000 
Mon Aug 20 17:49:13 2012 +0000 
Thu Apr 13 10:15:22 2017 -0500 
Mon Dec 11 23:30:20 2017 +0100 
\end{lstlisting}

\begin{lstlisting}[caption={Timestamps of Snippet-B}, label=lst:TS-B, firstnumber=1]
Thu Nov 6 09:04:08 2003 +0000
(*@\textbf{Mon Dec 13 19:12:36 2004 +0000}@*)
Mon Aug 20 17:49:13 2012 +0000
(*@\textbf{Wed Dec 21 16:08:28 2016 +0100}@*)
Thu Apr 13 10:15:22 2017 -0500
Mon Dec 11 23:30:20 2017 +0100
\end{lstlisting}
\end{minipage}

\section{Result}
\label{sec:result}

We will first present the results of Ant repository, followed by the analysis of all repositories.

\subsection{RQ1: How frequently are code clone snippets modified during development?}

\textbf{Ant Repository} ~~
Listing \ref{lst:Snippet-A} and \ref{lst:Snippet-B} show a code clone pair reported by the clone detector. These are type-2 clones, having different identifiers but the same structures and operators. Snippet-B includes a blank line and a comment line at the top because clone detectors generally tend to extend the clone areas as much as possible.

The timestamps in the commit logs for these two snippets are listed in Listings \ref{lst:TS-A} and \ref{lst:TS-B}. Lines 1, 2, 3, and 4 in Snippet-A correspond to lines 1, 3, 5, and 6 in Snippet-B, respectively, and those commit pairs are co-changed commits, verified by checking their commit hashes. Lines 2 and 4 in Snippet-B, listed in Listing \ref{lst:TS-B}, have no corresponding co-changed commits in Snippet-A. This indicates that only a single snippet of the clone pair was modified without the change of the other snippet at the same time. This could signify an unaligned change of clone pairs and a potential bad smell in the commit log of clones.

\begin{figure}[t]
\centerline{\includegraphics[width=9.0cm]{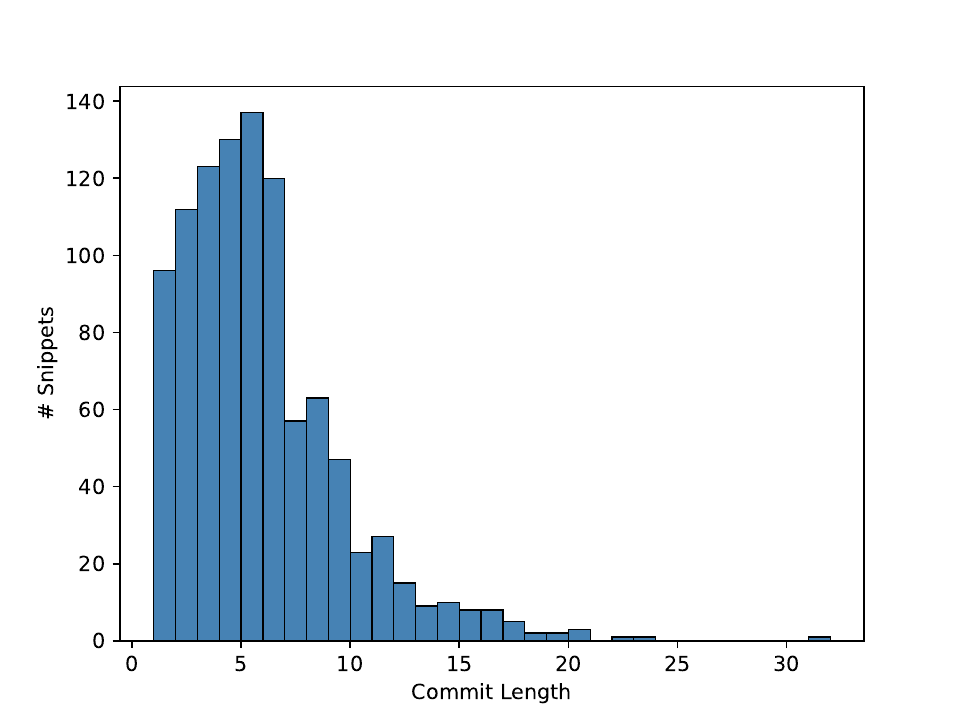}}
\caption{Commit Length of Clones in Ant}
\label{fig:ant-commit-len}
\end{figure}

Figure \ref{fig:ant-commit-len} shows a histogram of the length of the commit logs for each snippet of clone pairs in the Ant repository, representing the frequency of changes to the code clone snippets. The peak is around 5, with most changes being less than 10. The maximum length of the commit logs is 32, the minimum is 1, the average is 5.45, the median is 5, and the standard deviation is 3.68. This indicates that, on average, clone snippets are modified 5.45 times throughout the development process.

To determine if clone snippets are modified more or less frequently than random snippets (including clone and non-clone code), we examined the commit log lengths of 500 randomly selected code snippets in the Ant repository, each with a length of 19 lines, which matches the median length of Ant's clones. The average and the standard deviation of the commit log lengths of these random snippets are 5.50 and 3.77, respectively. A Welch's t-test indicates no statistical difference between the commit log lengths of clone snippets and random snippets. Thus, the change frequency of clone snippets has statistically no difference to that of the random snippets in the Ant repository.

We are interested in whether the two snippets of a clone pair have different commit lengths. For all clone pairs in the Ant repository, the differences in commit lengths have an average of 2.37 and a median of 1, indicating that the differences are generally small.

\textbf{All Repositories} ~~

\begin{figure}[t]
\centerline{\includegraphics[width=9.0cm]{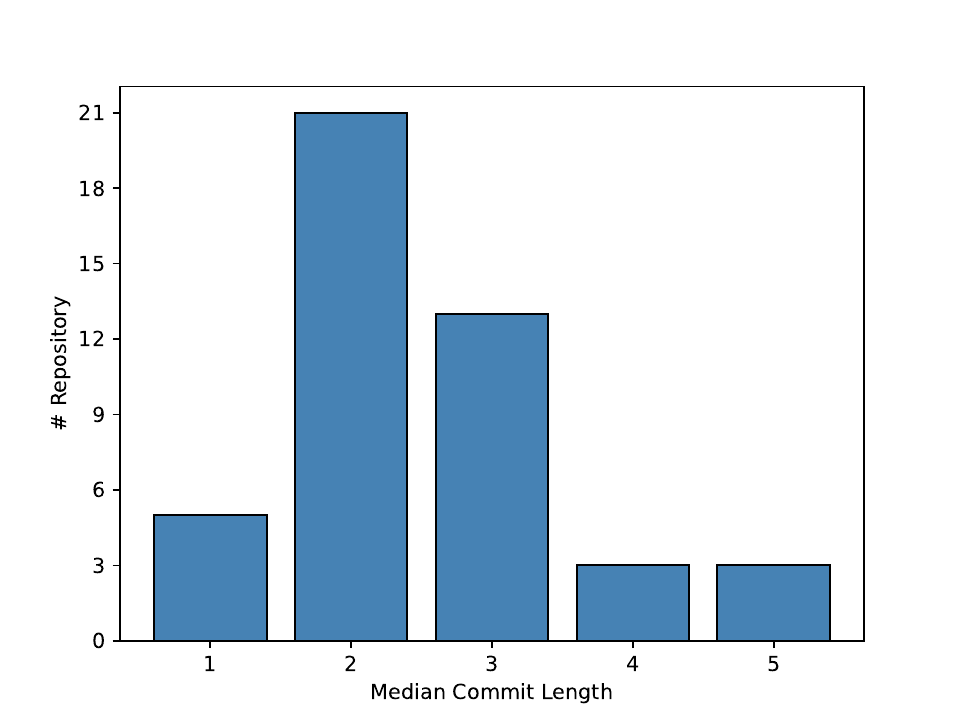}}
\caption{Median Commit Length of Clones in All Repositories}
\label{fig:commit-len-median-all}
\end{figure}

For all 45 repositories, we analyzed the commit log lengths of clones within each repository. Figure \ref{fig:commit-len-median-all} displays the median values of these commit log lengths. As shown in the graph, two commits are the most common across the repositories. The Ant repository, in particular, exhibits one of the highest numbers, with a median of 5 commits. Although we had anticipated a generally higher frequency of commits, this was not the case.

We investigated whether the commit log lengths, i.e., the changes to the code clone snippets are more or less frequent than those for other code snippets in the same repositories. For each repository, we randomly selected 500 code snippets whose code lengths matched the median length of the clone snippets in that repository. We then performed Welch's t-test to compare the two sets of commit log lengths. Among the 45 repositories analyzed, 20 repositories statistically showed that clone snippets are changed more frequently, 9 indicated that random snippets are changed more frequently, and 16 showed no statistically significant differences. Therefore, we cannot conclude that clone snippets are generally modified more or less frequently than other code snippets.

For the 29 cases (20+9 cases) with statistical differences, the discrepancy in the average frequency of changes was not substantial. For example, in the Tomee repository, clone snippets were changed 4.55 times on average, compared to 3.56 times for random snippets. 

\begin{screen}[4]
\textbf{Answer to RQ1: In most repositories, clones are generally stable, with the frequency of changes for clone snippets being low, typically only 2 or 3 times. The frequency of clone changes can be more, less, or the same as other code snippets, depending on the specific repository.}
\end{screen}


\subsection{RQ2:What proportion of commits are co-changed and are co-changed commits consistent?}
\label{sec:result-rq2}

\textbf{Ant Repository} ~~Ant had just 500 clone pairs; thus, there are 1,000 clone snippets in total. For these snippets, the total number of commits is 5,448. Among those, 2,476 commits have the same timestamps and hashes at the partner snippets, and they are co-changed commits. Therefore, the ratio of co-changed commits over all commits is 45.4\%.

We sought to determine whether co-changed commits are consistent. To rigorously ascertain this, it is necessary to expand and compare two versions of the source code before and after changes, ensuring that the modifications align well with each other and with their surrounding context. A consistency check of this nature requires a comprehensive understanding of the code and may sometimes necessitate labor-intensive human reviews. Therefore, we propose to detect potential inconsistencies using the patches (diffs) provided in the commit logs.

Listing \ref{lst:Patch} shows the patches 
in the latest commits of Snippet-A and Snippet-B from Listing \ref{lst:Snippet-A} and \ref{lst:Snippet-B}.
These snippets are similar and all aligned in the sense that two-line deletion and one-line addition at the corresponding lines.
We determined that they are well-aligned changes for the cloned snippets.

On the other hand, Listing \ref{lst:Patch-inconsistent} is an example of co-changed commits, and they contain many different comment lines. In addition, an important difference is a deletion of the top line of the latter snippet $this.out = out$, which cannot be seen in the former snippet. 
Thus, the consistency of this co-changed commit is not certain and a further inspection would be needed.
We refer to such uncertain co-changed commits that may cause potential inconsistencies as \emph{concerning (co-changed) commits} for a clone pair.

\begin{lstlisting}[float=tb, caption={Patches of Snippet-A and -B (Patch Similarity 0.907)}, label=lst:Patch, keywordstyle=\color{black}, numbers=none,breaklines=false]
Snippet-A's Patch
@@ -400,7 +400,6 @@
     if (u.children != null) {
(*@\textbf{-        List<UnknownElement> newC...}@*)
(*@\textbf{-        newChildren.addAll(u.children);}@*)
(*@\textbf{+        List<UnknownElement> newC...}@*)
         if (children != null) {
             newChildren.addAll(children);
         }
         children = newChildren;
--------------
Snippet-B's Patch
@@ -592,9 +592,8 @@
 
     // Children (this is a shadow of ...
     if (r.children != null) {
(*@\textbf{-        List<RuntimeConfigurable> newC...}@*)
(*@\textbf{-        newChildren.addAll(r.children);}@*)
(*@\textbf{+        List<RuntimeConfigurable> newC...}@*)
         if (children != null) {
             newChildren.addAll(children);
         }
         children = newChildren;
\end{lstlisting}


\begin{lstlisting}[float=tb,caption={Concerning Co-Changed Commits (Patch Similarity 0.09)}, label=lst:Patch-inconsistent, keywordstyle=\color{black}, numbers=none,breaklines=false]
98a63b... ant/taskdefs/ExecTask.java
@@ -186,7 +167,11 @@
     }
 
+    public void setInputString(String ...
+        redirector.setInputString(input...
+    }
+    
     /**
      * Controls whether error output of...
      * when output is being redirected...
      * Ant log
      */
--------
98a63b... ant/taskdefs/Java.java
@@ -305,1 +302,9 @@
(*@\textbf{-        this.out = out;}@*)
+    }
+
+    /**
+     * Set the input to use for the task
+     */
+    public void setInput(File input) {
+        redirector.setInput(input);
+    }
+
\end{lstlisting}

Based on the observation of patches, we have devised a method to infer the potential inconsistency of co-changed commits using the similarity of patches. We tokenized each patch, created word vectors, and measured their cosine similarity\footnote{We have used Python libraries \texttt{TfidfVectorizer} of \texttt{sklearn.feature\_extraction.text} and \texttt{cosine\_similarity} of \texttt{sklearn.metrics.pairwise}.} as \emph{patch similarity}. With this method, the patch similarity of Listing \ref{lst:Patch} is quite high at 0.907 (meaning the changes are very similar), whereas that of Listing \ref{lst:Patch-inconsistent} is significantly lower at 0.09 (meaning the changes are very different).

\begin{figure}[t]
\centerline{\includegraphics[width=9.0cm]{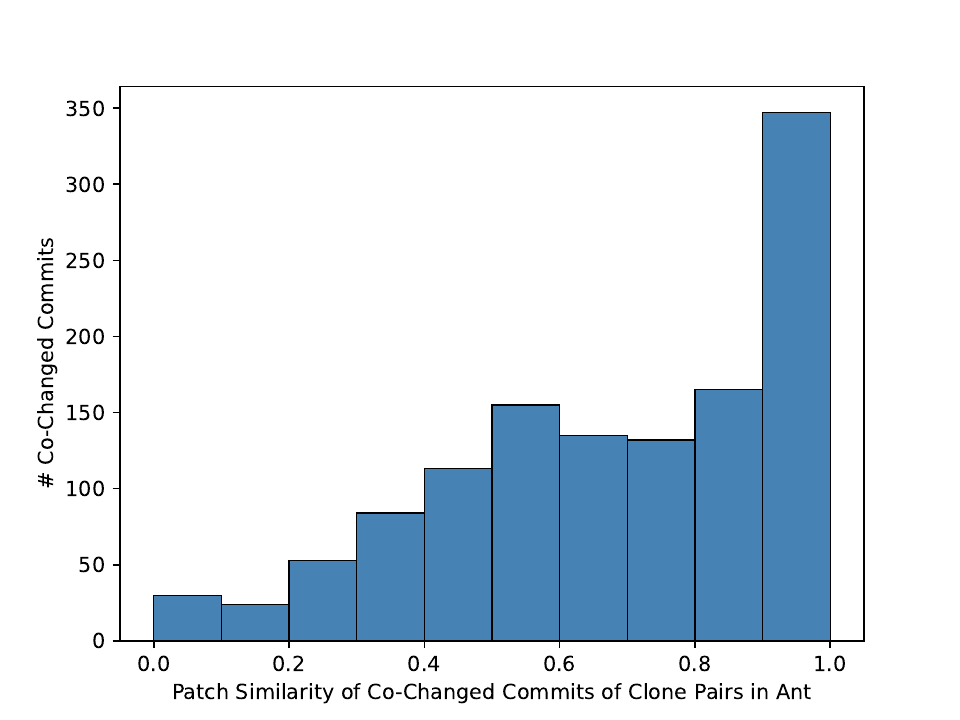}}
\caption{Patch Similarity Distribution of Co-Changed Commits in Ant}
\label{fig:patch-sim-cochanged-ant}
\end{figure}

Figure \ref{fig:patch-sim-cochanged-ant} shows the distribution of the patch similarity of the co-changed commits of clone pairs in Ant. The maximum,  minimum, average, and median values are 1.0, 0.01, 0.68, and 0.72, respectively.
As can be seen from this figure, while there are a small number of co-changed commits with values close to 0, there are many commits with values close to 1. 

After further analyzing and investigating various cases of patch differences and similarity values by the authors, we have set a threshold of 0.4 for potential inconsistency. Patches with a similarity value less than 0.4 are considered concerning co-changed commits for a clone pair in our analysis.

With this threshold value, the Ant repository contains 382 concerning co-changed commits which are 7.0\% of the total commits and 15.4\% of the co-changed ones.

\begin{figure}[t]
\centerline{\includegraphics[width=9.0cm]{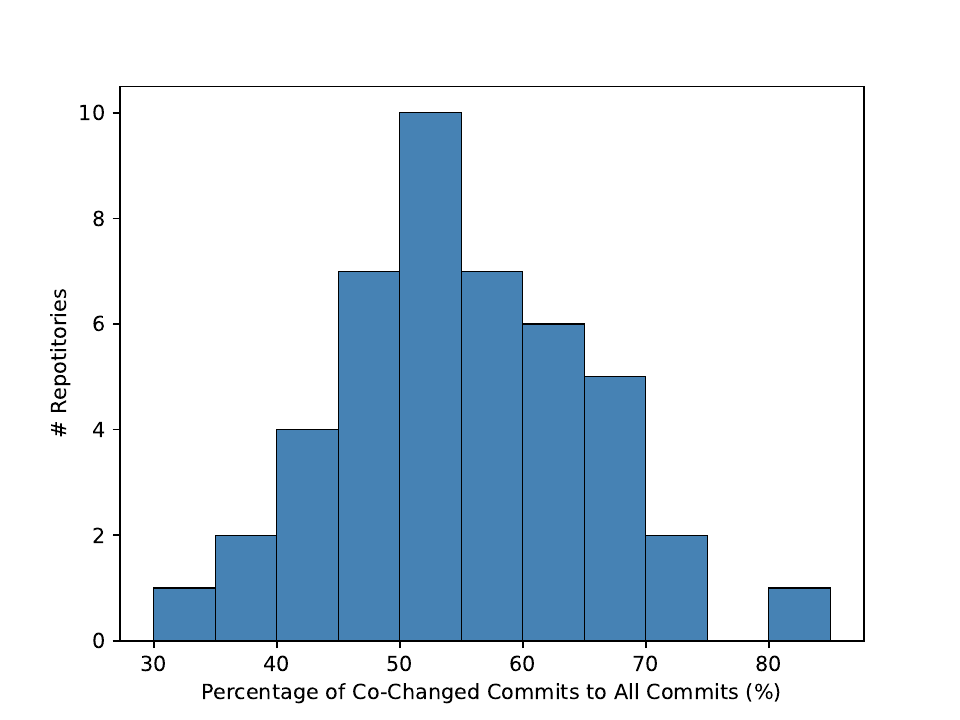}}
\caption{Co-changed Commit Ratio for All Repositories}
\label{fig:co-changed-ratio-all}
\end{figure}

\textbf{All Repositories} ~~We have investigated the ratio of co-changed commits over all commits for all 45 repositories. Figure \ref{fig:co-changed-ratio-all} shows the distribution of the co-changed commit ratio. In many repositories, we observe ratios between 40\% and 70\%, with the case of Ant falling within this range. The maximum ratio observed is 80.6\%, and the minimum is 31.4\%. The average ratio is 55.3\%, and the median is 53.5\%.
We can conclude that approximately half of the clone's commits are co-changed in each repository.

\begin{figure}[t]
\centerline{\includegraphics[width=9.0cm]{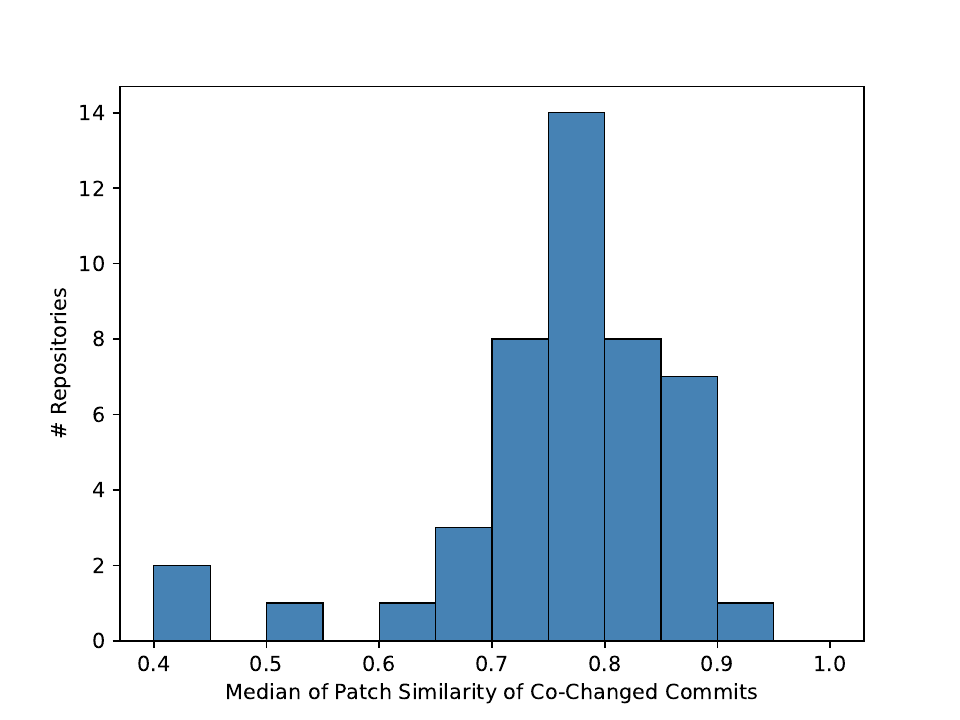}}
\caption{Patch Similarity Median of Co-Changed Commits for All Repositories}
\label{fig:patch-sim-cochanged-all}
\end{figure}

Patch similarity of the co-changed commits for all repositories has been investigated. Figure \ref{fig:patch-sim-cochanged-all} shows the distribution of the median values of each repository. We recognize that many repositories have a median over 0.65, while a few have less than 0.5, which would indicate the existence of some co-changed commits with the patch similarity less than the threshold.

\begin{figure}[t]
\centerline{\includegraphics[width=8.5cm]{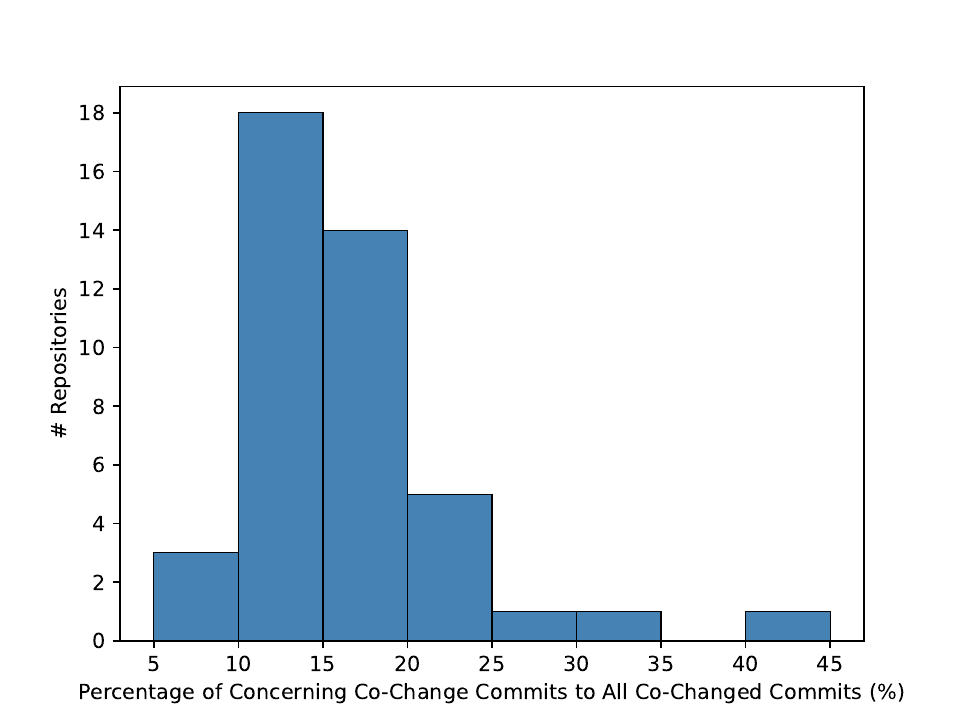}}
\caption{Ratio Distribution of Concerning co-changed commits over Co-Changed Commits in Each Repository}
\label{fig:concerning-over-cochange-all}
\end{figure}

We have thoroughly analyzed the ratios of the concerning co-changed commits over all co-changed commits for all repositories. Figure \ref{fig:concerning-over-cochange-all} presents their percentage over all co-changed commits. This figure means that in many repositories about 10\% to 20\% of co-changed commits are concerning ones.

\begin{screen}[4]
\textbf{Answer to RQ2: Across repositories, the ratio of co-changed commits is about half, and approximately 10-20\% of these are concerning co-changed commits, introducing potential inconsistency to clone snippets.}
\end{screen}

\subsection{RQ3: What are the concerning clone pairs that may need attention in the latest repository, and how prevalent are they?}

\begin{figure}[t]
\centerline{\includegraphics[width=9.0cm]{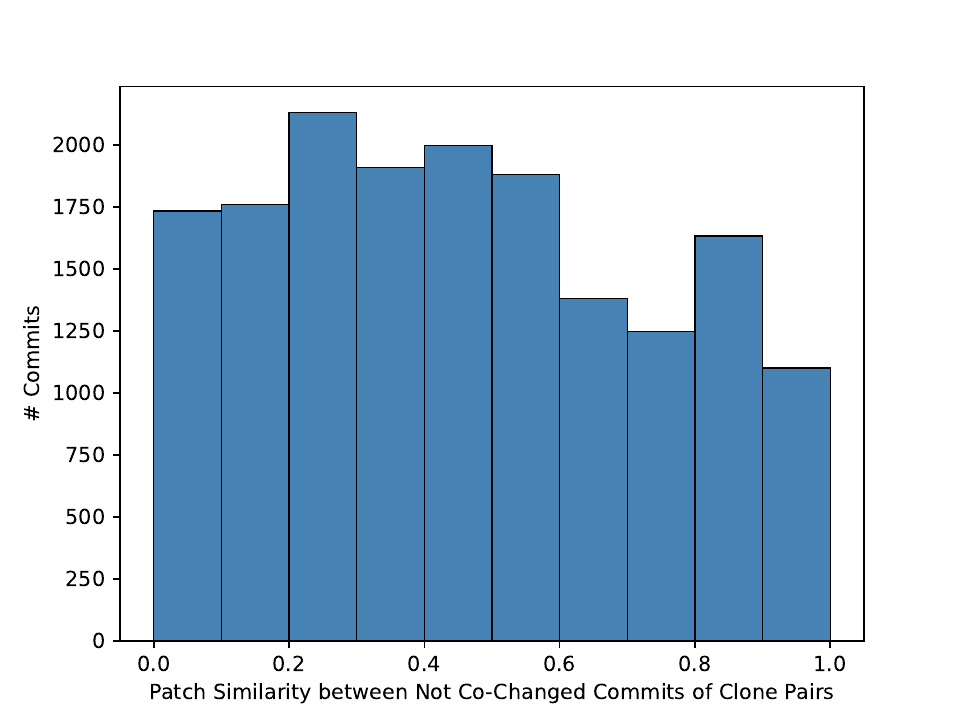}}
\caption{Patch Similarity Distribution of Not Co-Changed Commits in Ant}
\label{fig:patch-sim-notcochanged-ant}
\end{figure}

We have analyzed each commit in the commit logs of clone snippets, focusing on the similarity of patches and concerning co-changed commits. For the maintenance tasks of a current system, such as bug fixing and future refactoring, past concerning co-changed commits do not need to be addressed. Instead, attention should be given to the latest commits that form the current live code.

We are interested in whether the last commits of clone pairs are not co-changed in a single commit but are committed independently.
Figure \ref{fig:patch-sim-notcochanged-ant} illustrates the distribution of the patch similarity of not co-changed commits in Ant clone pairs. Unlike Figure \ref{fig:patch-sim-cochanged-ant}, which depicts the patch similarity of the co-changed commits, this figure shows a significant number of values close to zero, indicating low similarity between patches. Therefore, we infer that clone pairs not co-changed in the last commit tend to have low patch similarity, posing a risk of inconsistency.

In addition, we need to address the concerning co-changed commits discussed in RQ2. Here, we define two patterns of clone pairs, similar to the concerning co-changed commit. We refer to a clone pair exhibiting these patterns as a \emph{concerning clone pair}, which may require further attention for potential consistency.

\begin{itemize}
\item Pattern 1: Clone pairs whose last commits are not co-changed.
\item Pattern 2: Clone pairs whose last commits are co-changed but have patch similarity below the threshold (i.e., concerning co-changed commits).
\end{itemize}

\textbf{(Ant Repositories)} ~~We have investigated these patterns in the Ant repository. Among the 500 clone pairs in the Ant repository, 218 pairs (43.6\%) fall into Pattern 1, and 31 pairs (6.2\%) fall into Pattern 2. Therefore, a total of 249 pairs (49.8\%) of all clone pairs are identified as concerning.

\begin{figure}[t]
\centerline{\includegraphics[width=11.0cm]{
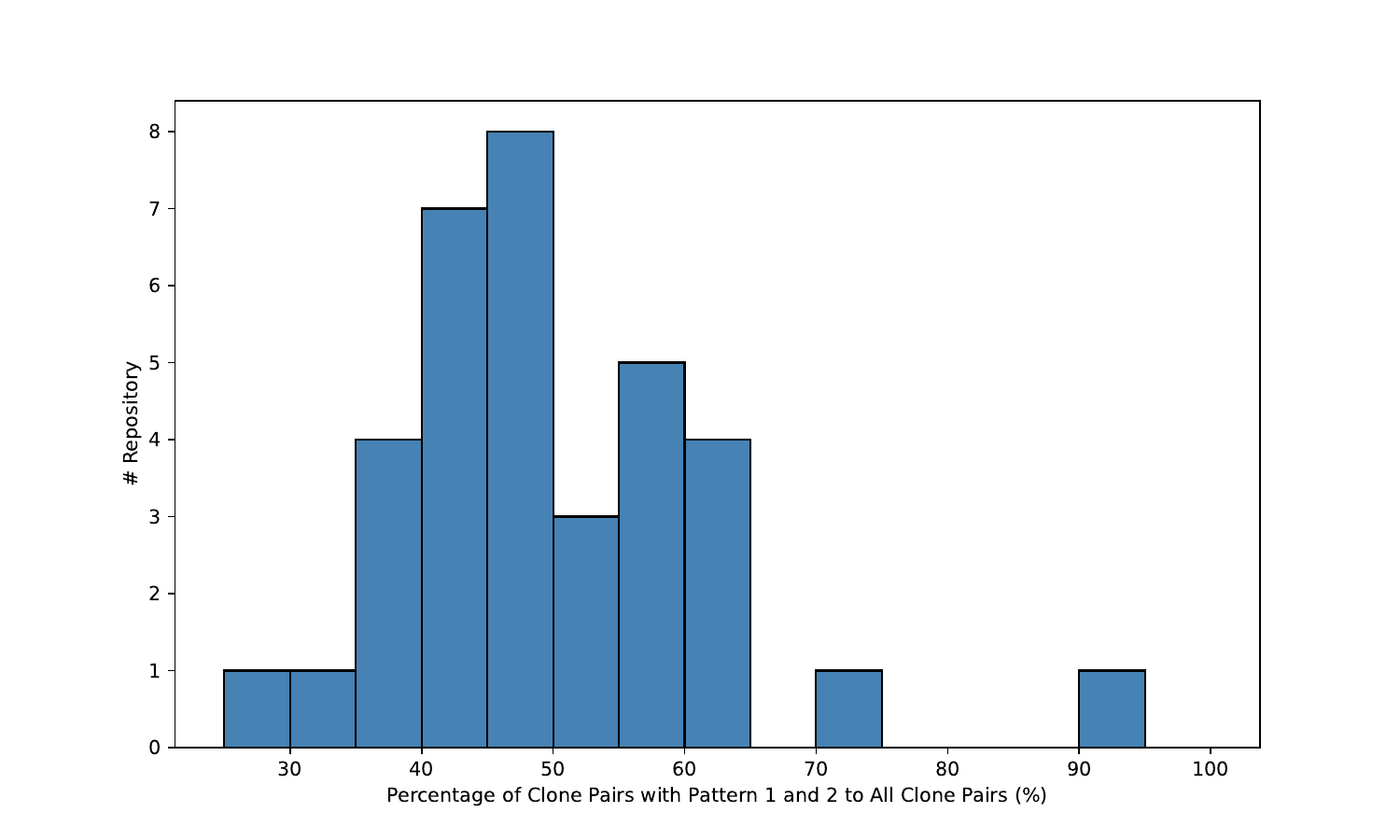}}
\caption{Ratio of Concerning Clone Pairs (Pattern 1 + 2) over All Clone Pairs for Each Repository}
\label{fig:pattern12-all}
\end{figure}

\textbf{(All Repositories)} ~~Figure \ref{fig:pattern12-all} illustrates the distribution of the ratio of the concerning clone pairs (Pattern 1 plus Pattern 2) over all clone pairs for each repository, indicating that 35\% to 65\% of clone pairs are concerning. Additionally, the ratios of Pattern 2 over Patterns 1 plus 2 were calculated, ranging from 0\% to 43.4\%, with an average of 13.7\%. This demonstrates that the majority of concerning clone pairs are due to independent last commits of clone pairs.

\begin{screen}[4]
\textbf{Answer to RQ3: Patterns of the concerning clone pairs in the latest repositories include those where the last commits either do not co-change or co-change with low patch similarity. In most repositories, these patterns account for 35\% to 65\% of all clone pairs.}
\end{screen}

\section{Discussions}
\subsection{Stability of Code Clone Snippets (RQ1)}

Initially, we hypothesized that code clones would be modified more frequently than other code snippets. This hypothesis was based on the expectation that code clone snippets would have logical coupling with their partner snippets, making them more susceptible to transitive changes from change demands randomly happening on source code. However, our results did not align with this hypothesis. One possible reason is that the logical coupling between code clone snippets is much weaker compared to the direct semantic coupling of program statements, and thus not strong enough to be statistically significant. Future research, involving larger-scale and more detailed analyses, may reveal the difference in the change frequency of clones and other snippets.

\subsection{Concerning co-changed commits (RQ2)}
\label{sec:rq2}

Bettenburg et al. reported that the percentage of inconsistent changes leading to software defects ranges between 1.26\% and 3.23\%\cite{betterburg09}. These values, derived from real bug history, are relatively low compared to our concerning commits and pairs. One reason for this difference could be their definition of `inconsistency' which means the change caused real fault occurrence. In such a sense, their analysis does not account for potential defects that are hidden or have not yet occurred. In contrast, our approach captures all cases of unaligned commits and low patch similarity as concerning commits and pairs. This might result in overestimation but ensures that potential future inconsistencies could be captured.


We have employed patch similarity as a measure of potential inconsistent change. This method has the advantage of easy data collection from Git and straightforward evaluation with a simple threshold. Since it does not account for the change details, it leaves room for potential misclassification. Accumulating data on past consistent and inconsistent changes of clones and applying machine learning methods would be one of the important future works.

Defining and determining consistent changes can be challenging, as consistency often depends on the context surrounding the code clones. For example, in type-2 clones where most variables undergo parameterized name changes, it might still be correct to leave one variable name unchanged, depending on the context. Using machine learning to learn the context around clones would help to determine consistency.

\subsection{Concerning Clone Pairs (RQ3)}
\label{sec:rq3}

The proportion of concerning clone pairs was notably high. One reason for this is that Pattern 1 includes all final commits that are not co-changed and are independent. As shown in Figure \ref{fig:patch-sim-notcochanged-ant}, even if commits of a pair are not co-changed, there may be cases to exhibit high patch similarity. Therefore, by examining the patch similarity of not co-changed commits and excluding those with high similarity, the proportion of concerning clone pairs might decrease. This will be one of our next research steps.

\subsection{Clone Sets with More Than Two Snippets}
\label{sec:morethan2}

To simplify the process and facilitate evaluation, the analysis was limited to clone sets with only two elements (clone snippets), i.e., clone pairs. Consequently, clone sets with three or more instances were excluded. Although the median size of clone sets in most repositories is 2 elements\footnote{In the 45 repositories, 40 have a median of 2 elements, and 5 have a median of 3 elements. The average number of elements ranges from 2.34 to 18.65.}, indicating that many clone sets are included in this analysis, those with three or more elements were overlooked. Analyzing sets with three or more elements by making combined pairs is straightforward, but the number of all combinations increases quadratically by $\binom{clone-set-size}{2}$ and could significantly impact the overall statistics. Additionally, defining `co-change' and `consistent' for sets with three or more elements becomes more complex. These issues will be addressed in future research.

\subsection{Implication for Code Clone Management System}
\label{sec:implecation}


The results of RQ1 indicate that the update frequency of code clones is not significantly high, nor is there a substantial difference compared to other codes. This raises questions about the cost-benefit of introducing comprehensive tools or IDEs solely for updating code clones. Instead, it seems desirable to add simpler warning features about the presence of code clones in the existing commit support features of various IDEs or editors, as well as alerts for cases where patch similarity is low, ensuring these features are seamlessly integrated with other features and easy to use.

If such warning features are implemented, users may receive many alerts according to the results of RQ2 and RQ3. However, as discussed in Sections \ref{sec:rq2} and \ref{sec:rq3}, the introduction of more accurate methods for determining consistency and filtering out false positives could mitigate this issue.


Numerous attempts have been reported to use machine learning for identifying code clones\cite{White2016DLC, wei2017supervised, Zhao2018DeepSim}. However, very few of these have been provided as practical systems. On the other hand, research on machine learning to detect and fix bugs in source codes is being actively pursued\cite{jian21,li20}, and practical systems such as GitHub copilot are actually in use\cite{copilot24}. As part of the training for such systems, it would be very interesting to train them to recognize code clones and their consistency, and to provide inconsistency alerts.

Many clones typically exist within a software system, including those created intentionally through copy-and-paste actions as well as those that are unintentional or accidental\cite{inoue21-2}. Additionally, clones can be categorized as either interesting or uninteresting. Therefore, the ability to filter and classify detected clones is essential and serves as the foundation for the features mentioned above.

\section{Threat to Validity}
\label{sec:threat}


\subsection{Internal Validity}
\paragraph{}
We used CCFinderSW as a code clone detection tool, but it has been noted that the results can vary depending on the tool used\cite{bellon07, roy09}. It is necessary to investigate what results might be obtained using other tools, such as NiCAD\cite{Cordy2011NiCad}, CCFinderX\cite{kamiya21}, or SourcererCC\cite{Sajnani2016SourcererCC}, in future studies.

\paragraph{}
We conducted our code clone detection using CCFinderSW with parameters fixed at minToken=50 (the
minimum number of tokens to detect), rnr=0.8 (the minimum
rate of non-repeating parts in the clone fragments),
and tks=12 (the minimum number of token types in the
clone fragments), but changing these parameters could significantly alter the results. For example, setting the rnr parameter to its default value of 0 may lead to the detection of more code clones, but this might also result in accidentally generated similar code being recognized as clones, potentially overshadowing the detection of intentionally copied and pasted code clones. 

\paragraph{}
In this study, we used the git-log to track the change history of code clone snippets. However, tracking changes with Git has its limitations\cite{ji20}. Changes in other parts of the code can make it impossible or erroneous to track clone snippets. Also, developers may alter Git's own logs through rebase or other operations. In our analysis, the commit history was sometimes confused due to rebasing, making it difficult to analyze, although this was very rare. Investigating more accurate methods for tracking the change history of code clones might be an intriguing topic.


\subsection{External Validity}

\paragraph{}
This analysis is based solely on the repository of 45 Java-based systems developed by the Apache project, and it may be difficult to directly apply these results to other open-source software (OSS) or projects in different languages. Understanding how differences in owners, contributors, and programming languages affect commit history is an interesting challenge for future research.

\paragraph{}
We conducted our statistical analysis with a fixed threshold for patch similarity set at 0.4, which is a criterion determined through observation of various cases within the Apache projects we studied. It is unclear whether this threshold can be generalized. It would be necessary to experimentally validate the criteria in other environments.

\section{Conclusion}

We have presented our analysis of 45 Apache projects aimed at characterizing the changes of code clones within these projects. 
Our findings indicate that, on average, clone snippets are modified infrequently, typically only two or three times throughout their lifetime. Also, the ratio of co-changed commits is about half, and 10-20\% of those are concerning ones, and the concerning clone pairs in the latest repositories account for 35-65\% of all clone pairs.

These results partially align with those of past empirical studies of code clones, although we have provided new insights necessary for the consistent and safe management of code clones.
As a new approach to code clone management, it is essential to incorporate a lightweight and highly compatible warning feature for code clones that activates during commit operations and integrates seamlessly with other features.

As part of our future work, we plan to extend the analysis of patches in the commit logs. We have employed a simple measure of similarity between two patches, which has room for improvement through more comprehensive methods, such as keyword prioritization/classification and statistical machine learning \cite{tufano18}. Additionally, we will continue to analyze other open source software and other programming languages as our targets, including clone sets with more than two elements.

\section*{Acknowledgment}

This work is supported by JSPS Grants-in-Aid for Scientific Research, Category (B), 23K28065, and Nanzan University Pache Research Subsidy I-A-2 for the 2024 academic year. 
We are grateful to Mr. Tatsuya Watanabe and Mr. Kyotaro Mase for their contributions to the clone data analysis.

\section*{Data Set Access}
For the replication and detailed validation of this empirical study, all data can be accessed through the following URL.\\
https://www.dropbox.com/scl/fo/00vh2erd9hateznwgb1ql\\
/AHMd-E8KakirIYIGb4YfLPA?rlkey=6ra0varc0cx6ac7v3\\
m3kccp6b\&e=1\&dl=0

\newpage


\bibliographystyle{IEEEtranS}

\bibliography{cloneCommit-bib}

\begin{thebibliography}{10}
\providecommand{\url}[1]{#1}
\csname url@samestyle\endcsname
\providecommand{\newblock}{\relax}
\providecommand{\bibinfo}[2]{#2}
\providecommand{\BIBentrySTDinterwordspacing}{\spaceskip=0pt\relax}
\providecommand{\BIBentryALTinterwordstretchfactor}{4}
\providecommand{\BIBentryALTinterwordspacing}{\spaceskip=\fontdimen2\font plus
\BIBentryALTinterwordstretchfactor\fontdimen3\font minus \fontdimen4\font\relax}
\providecommand{\BIBforeignlanguage}[2]{{%
\expandafter\ifx\csname l@#1\endcsname\relax
\typeout{** WARNING: IEEEtranS.bst: No hyphenation pattern has been}%
\typeout{** loaded for the language `#1'. Using the pattern for}%
\typeout{** the default language instead.}%
\else
\language=\csname l@#1\endcsname
\fi
#2}}
\providecommand{\BIBdecl}{\relax}
\BIBdecl

\bibitem{arshad22}
S.~Arshad, S.~Abid, and S.~Shamail, ``Codebert for code clone detection: A replication study,'' in \emph{2022 IEEE 16th International Workshop on Software Clones (IWSC)}, 2022, pp. 39--45.

\bibitem{Baker92aprogram}
B.~S. Baker, ``A program for identifying duplicated code,'' \emph{Proc. of Computing Science and Statistics: 24th Symposium on the Interface 24}, pp. 49--57, 1992.

\bibitem{baxter98}
I.~Baxter, A.~Yahin, L.~de~Moura, M.~Sant'Anna, and L.~Bier, ``Clone detection using abstract syntax trees,'' in \emph{Proc. of International Conference on Software Maintenance}, vol. 368-377, 01 1998, pp. 368--377.

\bibitem{bellon07}
S.~Bellon, R.~Koschke, G.~Antoniol, J.~Krinke, and E.~Merlo, ``Comparison and evaluation of clone detection tools,'' \emph{IEEE Transactions on Software Engineering}, vol.~33, no.~9, pp. 577--591, 2007.

\bibitem{betterburg09}
N.~Bettenburg, W.~Shang, W.~Ibrahim, B.~Adams, Y.~Zou, and A.~E. Hassan, ``An empirical study on inconsistent changes to code clones at release level,'' in \emph{Proceedings of the 2009 16th Working Conference on Reverse Engineering (WCRE09)}, 2009, pp. 85--94.

\bibitem{carter1993clone}
S.~Carter, R.~Frank, and D.~Tansley, ``Clone detection in telecommunications software systems: A neural net approach,'' in \emph{Proc. Int. Workshop on Application of Neural Networks to Telecommunications}, 1993, pp. 273--287.

\bibitem{Cordy2011NiCad}
J.~R. {Cordy} and C.~K. {Roy}, ``The nicad clone detector,'' in \emph{2011 IEEE 19th International Conference on Program Comprehension}, June 2011, pp. 219--220.

\bibitem{fowler99}
\BIBentryALTinterwordspacing
M.~Fowler, \emph{Refactoring - Improving the Design of Existing Code}, ser. Addison Wesley object technology series.\hskip 1em plus 0.5em minus 0.4em\relax Addison-Wesley, 1999. [Online]. Available: \url{http://martinfowler.com/books/refactoring.html}
\BIBentrySTDinterwordspacing

\bibitem{geiger06}
R.~Geiger, B.~Fluri, H.~C. Gall, and M.~Pinzger, ``Relation of code clones and change couplings,'' in \emph{Fundamental Approaches to Software Engineering}, L.~Baresi and R.~Heckel, Eds.\hskip 1em plus 0.5em minus 0.4em\relax Berlin, Heidelberg: Springer Berlin Heidelberg, 2006, pp. 411--425.

\bibitem{copilot24}
\BIBentryALTinterwordspacing
GitHub. (2024) Github copilot: The world’s most widely adopted ai developer tool. [Online]. Available: \url{https://github.com/features/copilot}
\BIBentrySTDinterwordspacing

\bibitem{goon17}
A.~Goon, Y.~Wu, M.~Matsushita, and K.~Inoue, ``Evolution of code clone ratios throughout development history of open-source c and c++ programs,'' \emph{International Workshop on Software Clones (IWSC2017)}, pp. 47--50, 2017.

\bibitem{higo04}
\BIBentryALTinterwordspacing
Y.~Higo, T.~Kamiya, S.~Kusumoto, and K.~Inoue, ``Refactoring support based on code clone analysis,'' in \emph{Product Focused Software Process Improvement, 5th International Conference, {PROFES} 2004, Kansai Science City, Japan, April}, ser. Lecture Notes in Computer Science, F.~Bomarius and H.~Iida, Eds., vol. 3009.\hskip 1em plus 0.5em minus 0.4em\relax Springer, 2004, pp. 220--233. [Online]. Available: \url{https://doi.org/10.1007/978-3-540-24659-6\_16}
\BIBentrySTDinterwordspacing

\bibitem{higo08}
\BIBentryALTinterwordspacing
Y.~Higo, S.~Kusumoto, and K.~Inoue, ``A metric-based approach to identifying refactoring opportunities for merging code clones in a java software system,'' \emph{J. Softw. Maintenance Res. Pract.}, vol.~20, no.~6, pp. 435--461, 2008. [Online]. Available: \url{https://doi.org/10.1002/smr.394}
\BIBentrySTDinterwordspacing

\bibitem{honda14}
A.~Honda, H.~Aman, T.~Sasaki, and M.~Kawahira, ``Investigation of convergence tendency of code clone ratio in open source development,'' \emph{IEICE Transactions D}, vol. J97-D, no.~7, pp. 1213--1215, 2014.

\bibitem{hotta10}
K.~Hotta, Y.~Sano, Y.~Higo, and S.~Kusumoto, ``Is duplicate code more frequently modified than non-duplicate code in software evolution? an empirical study on open source software,'' in \emph{Proceedings of IWPSE-EVOL2010}, ser. IWPSE-EVOL '10.\hskip 1em plus 0.5em minus 0.4em\relax New York, NY, USA: ACM, 2010, p. 73–82.

\bibitem{inoue21-2}
K.~Inoue, ``Introduction to code clone analysis,'' in \emph{Code Clone Analysis}, K.~Inoue and C.~K. Roy, Eds.\hskip 1em plus 0.5em minus 0.4em\relax Springer Singapore, 2021, pp. 3--27.

\bibitem{inoue21}
K.~{Inoue} and C.~K. {Roy}, Eds., \emph{Code Clone Analysis, Research, Tools and Practices}.\hskip 1em plus 0.5em minus 0.4em\relax Springer, 2021.

\bibitem{islam16}
J.~F. Islam, M.~Mondal, and C.~K. Roy, ``Bug replication in code clones: An empirical study,'' in \emph{2016 IEEE 23rd International Conference on Software Analysis, Evolution, and Reengineering (SANER)}, vol.~1, 2016, pp. 68--78.

\bibitem{ji20}
T.~Ji, L.~Chen, X.~Yi, and X.~Mao, ``Understanding merge conflicts and resolutions in git rebases,'' in \emph{2020 IEEE 31st International Symposium on Software Reliability Engineering (ISSRE)}, 2020, pp. 70--80.

\bibitem{jian21}
N.~Jiang, T.~Lutellier, and L.~Tan, ``Cure: Code-aware neural machine translation for automatic program repair,'' in \emph{2021 IEEE/ACM 43rd International Conference on Software Engineering (ICSE)}, 2021, pp. 1161--1173.

\bibitem{kamiya21}
T.~Kamiya, ``Ccfinderx: An interactive code clone analysis environment,'' in \emph{Code Clone Analysis}, K.~Inoue and C.~K. Roy, Eds.\hskip 1em plus 0.5em minus 0.4em\relax Springer, 2021, pp. 31--44.

\bibitem{Kamiya2002CCFinder}
T.~Kamiya, S.~Kusumoto, and K.~Inoue, ``Ccfinder: A multilinguistic token-based code clone detection system for large scale source code,'' \emph{IEEE Trans. Software Eng.}, vol.~28, pp. 654--670, 2002.

\bibitem{kapser06}
C.~Kapser and M.~W. Godfrey, ``"cloning considered harmful" considered harmful,'' in \emph{2006 13th Working Conference on Reverse Engineering}, 2006, pp. 19--28.

\bibitem{kim05}
M.~Kim, V.~Sazawal, D.~Notkin, and G.~Murphy, ``An empirical study of code clone genealogies,'' \emph{ACM SIGSOFT Software Engineering Notes}, vol.~30, no.~5, pp. 187--196, 2005.

\bibitem{koschke07}
R.~Koschke, ``Survey of research on software clones,'' in \emph{Dagstuhl Seminar Proceedings}.\hskip 1em plus 0.5em minus 0.4em\relax Schloss Dagstuhl-Leibniz-Zentrum f{\"u}r Informatik, 2007.

\bibitem{jens07}
J.~Krinke, ``A study of consistent and inconsistent changes to code clones,'' in \emph{14th Working Conference on Reverse Engineering (WCRE 2007)}, 2007, pp. 170--178.

\bibitem{li20}
\BIBentryALTinterwordspacing
Y.~Li, S.~Wang, and T.~N. Nguyen, ``Dlfix: context-based code transformation learning for automated program repair,'' in \emph{Proceedings of the ACM/IEEE 42nd International Conference on Software Engineering}, ser. ICSE '20.\hskip 1em plus 0.5em minus 0.4em\relax New York, NY, USA: Association for Computing Machinery, 2020, p. 602–614. [Online]. Available: \url{https://doi.org/10.1145/3377811.3380345}
\BIBentrySTDinterwordspacing

\bibitem{lozano14}
A.~Lozano, F.~Jaafar, K.~Mens, and Y.-G. Guéhéneuc, ``Clones and macro co-changes,'' in \emph{Proceedings of the 8th International Conference on Software Clones (IWSC2014)}, 2014, pp. 1--14.

\bibitem{marks13}
L.~Marks, Y.~Zou, and I.~Keivanloo, ``An empirical study of the factors affecting co-change frequency of cloned code,'' in \emph{Proceedings of the 2013 Conference of the Center for Advanced Studies on Collaborative Research}, ser. CASCON '13.\hskip 1em plus 0.5em minus 0.4em\relax USA: IBM Corp., 2013, p. 161–175.

\bibitem{mondal19}
M.~Mondal, B.~Roy, C.~K. Roy, and K.~A. Schneider, ``Ranking co-change candidates of micro-clones,'' in \emph{Proceedings of CASCON '19}.\hskip 1em plus 0.5em minus 0.4em\relax USA: IBM Corp., 2019, p. 244–253.

\bibitem{mondal13}
M.~Mondal, C.~K. Roy, and K.~A. Schneider, ``Insight into a method co-change pattern to identify highly coupled methods: An empirical study,'' in \emph{2013 21st International Conference on Program Comprehension (ICPC)}, 2013, pp. 103--112.

\bibitem{mondal14}
------, ``A fine-grained analysis on the evolutionary coupling of cloned code,'' in \emph{2014 IEEE International Conference on Software Maintenance and Evolution}, 2014, pp. 51--60.

\bibitem{mondal14-2}
\BIBentryALTinterwordspacing
------, ``Prediction and ranking of co-change candidates for clones,'' in \emph{Proceedings of the 11th Working Conference on Mining Software Repositories}, ser. MSR 2014.\hskip 1em plus 0.5em minus 0.4em\relax New York, NY, USA: Association for Computing Machinery, 2014, p. 32–41. [Online]. Available: \url{https://doi.org/10.1145/2597073.2597104}
\BIBentrySTDinterwordspacing

\bibitem{mondal20}
\BIBentryALTinterwordspacing
------, ``A survey on clone refactoring and tracking,'' \emph{Journal of Systems and Software}, vol. 159, p. 110429, 2020. [Online]. Available: \url{https://www.sciencedirect.com/science/article/pii/S0164121219302031}
\BIBentrySTDinterwordspacing

\bibitem{mostafa19}
M.~J.~I. Mostafa, ``An empirical study on clone evolution by analyzing clone lifetime,'' in \emph{2019 IEEE 13th International Workshop on Software Clones (IWSC)}, 2019, pp. 20--26.

\bibitem{yudha12}
H.~A. Myrizki Sandhi~Yudha, Ryohei~Asano, ``A feature analysis of co-changed code clone by using clone metrics,'' \emph{IEICE TRANSACTIONS on Fundamentals of Electronics, Communications and Computer Sciences}, vol. E95-A, no.~9, pp. 1498--1500, 2012.

\bibitem{rabbani22}
S.~M. Rabbani, N.~Ahmad~Gulzar, S.~Arshad, S.~Abid, and S.~Shamail, ``A comparative analysis of clone detection techniques on semanticclonebench,'' in \emph{2022 IEEE 16th International Workshop on Software Clones (IWSC)}, 2022, pp. 16--22.

\bibitem{rahman17}
M.~S. Rahman and C.~K. Roy, ``On the relationships between stability and bug-proneness of code clones: An empirical study,'' in \emph{2017 IEEE 17th International Working Conference on Source Code Analysis and Manipulation (SCAM)}, 2017, pp. 131--140.

\bibitem{rattan13}
D.~Rattan, R.~Bhatia, and M.~Singh, ``Software clone detection: A systematic review,'' \emph{Information and Software Technology}, vol.~55, no.~7, pp. 1165--1199, 2013.

\bibitem{roy08}
C.~K. Roy and J.~R. Cordy, ``An empirical study of function clones in open source software,'' in \emph{2008 15th Working Conference on Reverse Engineering}, 2008, pp. 81--90.

\bibitem{roy09}
\BIBentryALTinterwordspacing
C.~K. Roy, J.~R. Cordy, and R.~Koschke, ``Comparison and evaluation of code clone detection techniques and tools: A qualitative approach,'' \emph{Science of Computer Programming}, vol.~74, no.~7, pp. 470 -- 495, 2009. [Online]. Available: \url{http://www.sciencedirect.com/science/article/pii/S0167642309000367}
\BIBentrySTDinterwordspacing

\bibitem{roy07}
C.~K. Roy and J.~R. Cordy, ``A survey on software clone detection research,'' \emph{Queen’s School of Computing TR}, vol. 541, no. 115, pp. 64--68, 2007.

\bibitem{ripon10}
R.~K. Saha, M.~Asaduzzaman, M.~F. Zibran, C.~K. Roy, and K.~A. Schneider, ``Evaluating code clone genealogies at release level: An empirical study,'' in \emph{2010 10th IEEE Working Conference on Source Code Analysis and Manipulation}, 2010, pp. 87--96.

\bibitem{OREO-FSE2018}
V.~Saini, F.~Farmahinifarahani, Y.~Lu, P.~Baldi, and C.~V. Lopes, ``Oreo: Detection of clones in the twilight zone,'' in \emph{Proc. 2018 {ACM} Joint Meeting on European Software Engineering Conference / Symposium on the Foundations of Software Engineering, {ESEC/SIGSOFT} {FSE} 2018, Lake Buena Vista, FL, Nov.}, 2018, pp. 354--365.

\bibitem{Sajnani2016SourcererCC}
\BIBentryALTinterwordspacing
H.~Sajnani, V.~Saini, J.~Svajlenko, C.~K. Roy, and C.~V. Lopes, ``Sourcerercc: Scaling code clone detection to big-code,'' in \emph{Proceedings of the 38th International Conference on Software Engineering}, ser. ICSE '16.\hskip 1em plus 0.5em minus 0.4em\relax New York, NY, USA: ACM, 2016, pp. 1157--1168. [Online]. Available: \url{http://doi.acm.org/10.1145/2884781.2884877}
\BIBentrySTDinterwordspacing

\bibitem{semura17}
Y.~Semura, N.~Yoshida, E.~Choi, and K.~Inoue, ``Ccfindersw: Clone detection tool with flexible multilingual tokenization,'' in \emph{Proceedings of the 24th Asia-Pacific Software Engineering Conference (APSEC 2017), Nanjing, China}, 2017, pp. 654--659.

\bibitem{tufano18}
\BIBentryALTinterwordspacing
M.~Tufano, C.~Watson, G.~Bavota, M.~Di~Penta, M.~White, and D.~Poshyvanyk, ``Deep learning similarities from different representations of source code,'' in \emph{Proceedings of the 15th International Conference on Mining Software Repositories}, ser. MSR '18.\hskip 1em plus 0.5em minus 0.4em\relax New York, NY, USA: Association for Computing Machinery, 2018, p. 542–553. [Online]. Available: \url{https://doi.org/10.1145/3196398.3196431}
\BIBentrySTDinterwordspacing

\bibitem{bladel23}
B.~van Bladel and S.~Demeyer, ``A comparative study of code clone genealogies in test code and production code,'' in \emph{2023 IEEE International Conference on Software Analysis, Evolution and Reengineering (SANER)}, 2023, pp. 913--920.

\bibitem{wei2017supervised}
H.~Wei and M.~Li, ``Supervised deep features for software functional clone detection by exploiting lexical and syntactical information in source code,'' in \emph{IJCAI}, 2017, pp. 3034--3040.

\bibitem{White2016DLC}
\BIBentryALTinterwordspacing
M.~White, M.~Tufano, C.~Vendome, and D.~Poshyvanyk, ``Deep learning code fragments for code clone detection,'' in \emph{Proceedings of the 31st IEEE/ACM International Conference on Automated Software Engineering}, ser. ASE 2016.\hskip 1em plus 0.5em minus 0.4em\relax New York, NY, USA: ACM, 2016, pp. 87--98. [Online]. Available: \url{http://doi.acm.org/10.1145/2970276.2970326}
\BIBentrySTDinterwordspacing

\bibitem{xie14}
S.~Xie, F.~Khomh, Y.~Zou, and I.~Keivanloo, ``An empirical study on the fault-proneness of clone migration in clone genealogies,'' in \emph{2014 Software Evolution Week - IEEE Conference on Software Maintenance, Reengineering, and Reverse Engineering (CSMR-WCRE)}, 2014, pp. 94--103.

\bibitem{Zhao2018DeepSim}
G.~Zhao and J.~Huang, ``Deepsim: Deep learning code functional similarity,'' in \emph{Proceedings of the 2018 26th ACM Joint Meeting on European Software Engineering Conference and Symposium on the Foundations of Software Engineering}, ser. ESEC/FSE 2018.\hskip 1em plus 0.5em minus 0.4em\relax New York, NY, USA: ACM, 2018, pp. 141--151.

\end{thebibliography}

\end{document}